\documentclass[aps, twocolumn, superscriptaddress, prb]{revtex4-1}
\usepackage{graphicx}% Include figure files
\usepackage{dcolumn}% Align table columns on decimal point
\usepackage{bm}% bold math
\usepackage{pifont}
\usepackage{amsmath}
\usepackage{tabularx}
\usepackage{color}
\usepackage{float}
\usepackage{booktabs}
\usepackage{amsmath}
\usepackage{amssymb}
\usepackage[utf8]{inputenc}
\usepackage{amsmath}
\usepackage{subfigure}

\begin{document}
\preprint{APS/123-QED}

\title{Highly anisotropic strain dependencies in PrIr$_2$Zn$_{20}$}

\author{A. Wörl}
\affiliation{Experimentalphysics VI, Center for Electronic Correlations and Magnetism, Augsburg University, Germany}
\author{T. Onimaru}
\affiliation{Graduate School of Advanced Sciences of Matter, Hiroshima University, Japan}
\author{Y. Tokiwa}
\affiliation{Experimentalphysics VI, Center for Electronic Correlations and Magnetism, Augsburg University, Germany}
\author{Y. Yamane}
\affiliation{Graduate School of Advanced Sciences of Matter, Hiroshima University, Japan}
\author{K. T. Matsumoto}
\affiliation{Graduate School of Advanced Sciences of Matter, Hiroshima University, Japan}
\author{T. Takabatake}
\affiliation{Graduate School of Advanced Sciences of Matter, Hiroshima University, Japan}
\author{P. Gegenwart}
\affiliation{Experimentalphysics VI, Center for Electronic Correlations and Magnetism, Augsburg University, Germany}
\date{\today}% It is always \today, today,

\begin{abstract}
We report thermal expansion and magnetostriction of the cubic non-Kramers system PrIr$_2$Zn$_{20}$ with a non-magnetic $\varGamma_{3}$ ground state doublet. In previous experiments, antiferroquadrupolar order at \hbox{$T_{\mathrm{Q}}=0.11$\,K} and a Fermi liquid state around $B_{\mathrm{c}}\approx5$\,T for \hbox{$\boldsymbol{B}\parallel[001]$}, indicative of possible ferrohastatic order, were discovered. For magnetic fields \hbox{$\boldsymbol{B}\parallel[001]$}, the low temperature longitudinal and transverse thermal expansion and magnetostriction are highly anisotropic. The resulting volume strain is very small, indicating that the Pr valence remains nearly constant as a function of magnetic field. We conclude that the Fermi liquid state around $B_{\mathrm{c}}$  forms through a very little change in c-f hybridization. This result is in sharp contrast to Ce- and Yb-based Kramers Kondo lattices which show significantly larger volume strains due to the high sensitivity of the Kondo temperature to hydrostatic pressure.
\end{abstract}

\maketitle
In recent years, comprehensive studies on heavy fermion (HF) materials imparted fundamental understanding on their competing ground states which are classified as strongly hybridized, magnetically ordered and in special cases quantum critical \citep{Gegenwart08}. As proposed by Doniach in 1977,  the ground state depends on the interaction strength $J$ between localized magnetic moments and conduction electrons which can be tuned by the variation of an external control parameter, e.g. magnetic field, pressure or chemical substitution \cite{Doniach77}. Extensive research on materials with quadrupolar degrees of freedom followed, in order to verify the applicability of the Doniach picture. Even though a generic phase diagram has not been established yet, a variety of novel quadrupole driven states were detected, such as exotic antiferroquadrupolar (AFQ) order in PrPb$_3$ \cite{Onimaru05}, HF superconductivity in PrOs$_4$Sb$_{12}$ \cite{Bauer02} and PrV$_2$Al$_{20}$ \cite{Tsujimoto14} and signatures of two-channel Kondo effect in PrIr$_{2}$Zn$_{20}$ \cite{Onimaru16}. \par In particular, the material class of cubic Pr-based 1-2-20 systems, with the non-Kramers $\varGamma_3$ ground state doublet, provides key prerequisites to explore purely orbital driven physics \cite{Sakai11, Onimaru16S}. Considerable efforts have been expended on characterizing the materials PrIr$_{2}$Zn$_{20}$, PrRh$_{2}$Zn$_{20}$, PrV$_2$Al$_{20}$ and PrTi$_2$Al$_{20}$, which share the coexistence of quadrupolar order and superconductivity \citep{Onimaru11, Onimaru12, Tsujimoto14, Sakai12}. The high coordination number and the local $T_d$ symmetry of the Pr-ions facilitate the hybridization of electric quadrupole moments and conduction electrons, whereby thermopower measurements suggest enhanced hybridization effects for the Al-based systems as compared to the Zn-based systems \cite{Machida2015}. 

Up to now it remains elusive, whether a quadrupolar quantum critical state, driven by strong correlations between the fluctuating order parameter and conduction electrons, can evolve in those systems. First indications could be found for PrTi$_2$Al$_{20}$\citep{Matsu2012}, where the application of hydrostatic pressure significantly enhances the superconducting transition temperature from $T_{\mathrm{c}}=0.2$\,K ($p=0$) to 1.1\,K ($p=8.7$\,GPa) as well as the effective mass from $m^{*}/m\approx16$ to around $110$ .  Further hydrostatic pressure suppresses the ferroquadrupolar order. To reveal universal characteristics of the quadrupolar Kondo lattice materials, systematic studies in magnetic field and under hydrostatic/uniaxial pressure are necessary.
  
In this work we focus on the quadrupolar Kondo lattice PrIr$_2$Zn$_{20}$, which crystallizes in the cubic CeCr$_2$Al$_{20}$-type structure with $Fd\bar{3}m$ space group \citep{Nasch97}. The crystalline electric field (CEF) ground state is the non-Kramers $\varGamma_3$ doublet, which carries two quadrupoles ($O_{2}^{0}$, $O_{2}^{2}$) and one octupole ($T_{xyz}$). The energy gap between the ground state and the first excited  $\varGamma_4$ triplet state is $\Delta_{\mathrm{CEF}}=28$\,K \citep{Onimaru16S}. At low temperatures, non-Fermi liquid behavior, a key signature of the two-channel Kondo effect, was observed \citep{Onimaru16}. Origin of this "strange" metallic state is the overscreening of quadrupole moments by spin up and down conduction electron bands. Theoretically, the two-channel Kondo ground state is associated with a residual entropy \cite{Cox98} of $S=0.5R\ln2$. In the case of PrIr$_2$Zn$_{20}$ this entropy is released by AFQ order at \hbox{$T_{\mathrm{Q}}=0.11$\,K}. When applying a magnetic field \hbox{$B_{\mathrm{c}}\approx 5$\,T} along the [001] direction \cite{Onimaru11}, the AFQ order is suppressed. In vicinity of $B_{\mathrm{c}}$, pronounced anomalies in Seebeck coefficient  \cite{Ikeura14}, specific heat\cite{Onimaru11} and elastic constants \cite{Ishii11} as well as a peculiar Fermi liquid state in the electrical resistivity \citep{Onimaru16} were observed. An explanation for those phenomena might be a field induced ferrohastatic order, where localized 4f$^2$-moments hybridize exclusively with the spin up conduction electron band, forming a Fermi liquid with a small hybridization gap \citep{Hoshino11, VD2018}. Since hastatic order involves a spinorial hybridization, which breaks double time reversal symmetry, it is distinctly different from the Kondo hybridization in Kramers materials. The concept of the hastatic order \cite{Chandra13} was originally introduced to explain the low temperature "hidden order" phase of the tetragonal material URu$_2$Si$_2$.

In order to trace the evolution of hybridization in PrIr$_2$Zn$_{20}$ as a function of magnetic field and temperature, we use the thermodynamic properties volume thermal expansion and magnetostriction. In general, hybridization and rare-earth ion valence are closely related properties, since an alteration of hybridization leads to a valence change. On the other hand, the valence of the rare-earth ion scales with its volume. Therefore, volume thermal expansion and magnetostriction are very suitable probes to detect changes in hybridization.

To perform the thermal expansion and magnetostriction measurements, we utilized a dilution refrigerator equipped with a 13\,T superconducting magnet. Linear thermal expansion \hbox{$\alpha=1/L\,(\mathrm{d}\Delta L/\mathrm{d}T)$} and magnetostriction $\lambda=1/L\,(\mathrm{d}\Delta L/\mathrm{d}B)$, where $\Delta L$ denotes the relative length change and $L$ the sample length at room temperature, were measured by use of a miniaturized capacitance dilatometer made of copper beryllium \cite{Kuechler17}. %The small dimensions allowed to rotate the dilatometer in the bore of the superconducting magnet and enabled the measurement of $\alpha$ and $\lambda$ perpendicular to magnetic field.
To deduce the relative length change from the measured capacitance value, we applied the Pott and Schefzyk principle which takes account of the maximal adjustable capacitance of the dilatometer \cite{Pott83}. Linear thermal expansion coefficient $\alpha$ and magnetostriction coefficient $\lambda$ were determined by numerical differentiation of the relative length change with respect to temperature and magnetic field. %For the calculation of $\lambda$ a differentiation window of $\Delta B=0.05$\,T was used. To determine $\alpha$, the differentiation window was set to $\Delta T=0.015$\,K at the lowest temperatures and gradually increased to $0.08$\,K at the highest measured temperature of $T=5$\,K. 
The investigated single crystalline sample was grown by a Zn self-flux technique, as described by Saiga \textit{et al.} \citep{Saiga08}, with a length of $L=1.295$\,mm along the [001] direction and a residual resistivity ratio of $RRR=54$. 

\begin{figure}[htbp]
\includegraphics[width=\linewidth,keepaspectratio]{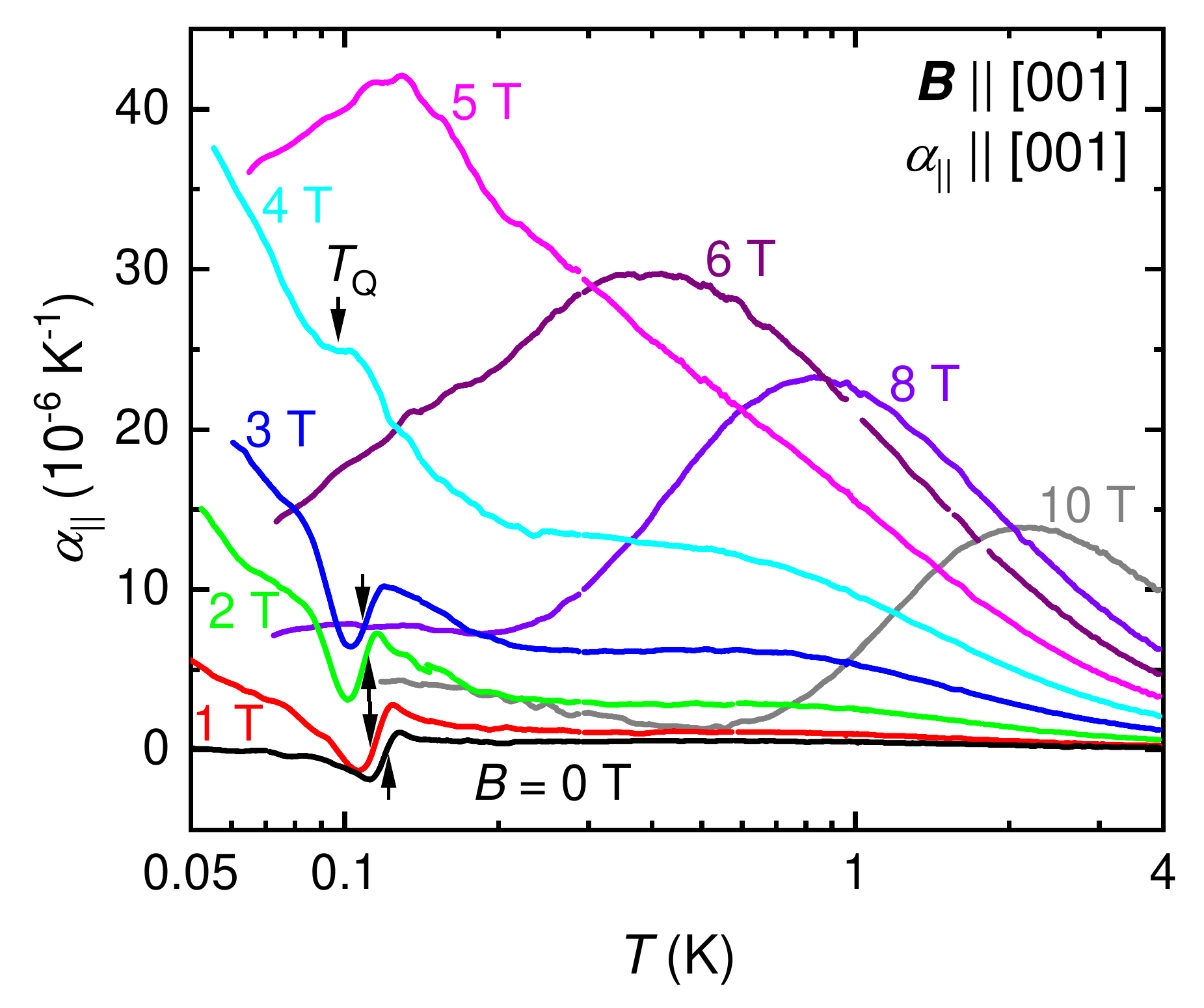}
\caption{Temperature dependence of the longitudinal thermal expansion coefficient ($\alpha_{\parallel}$) for magnetic fields $\boldsymbol{B}\parallel[001]$. The onset of antiferroquadrupolar order is marked by arrows.}
\label{fig:TEsingle}
\end{figure}

First, we present and discuss the results of the thermal expansion measurements. Fig. \ref{fig:TEsingle} shows the temperature dependence of the longitudinal thermal expansion coefficient ($\alpha_{\parallel}$) for $\boldsymbol{B} \parallel [001]$. 
\begin{figure}[htbp]
\includegraphics[width=\linewidth,keepaspectratio]{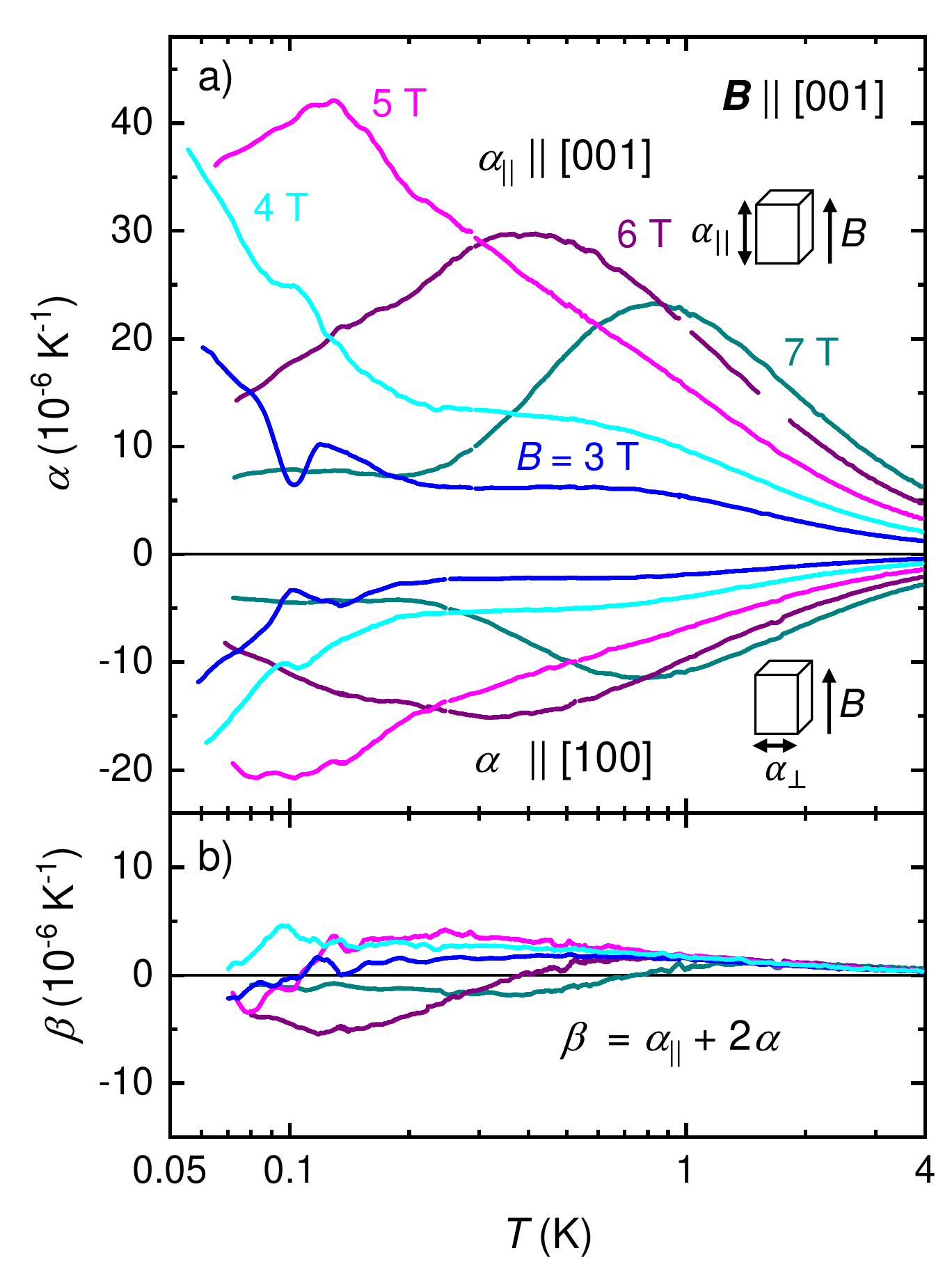}
\caption{(a) Temperature dependence of the longitudinal and transverse thermal expansion coefficients ($\alpha_{\parallel}$ and $\alpha_{\perp}$) for magnetic fields $\boldsymbol{B}\parallel [001]$ , with $3\,\mathrm{T} \le B \le 7\, \mathrm{T}$. (b) Temperature dependence of the volume thermal expansion coefficient ($\beta=a_{\parallel }+2\alpha_{\perp}$).}% (b) Illustration of the simulated thermal expansion coefficients $\alpha_{\parallel}$ and $\alpha_{\perp}$ at different magnetic fields $\boldsymbol{B}\parallel [001]$. The inset shows direct comparison between experiment (color) and calculation (black) for $\alpha_{\perp}$. \textcolor{blue}{Simulation results for $B\le4$\,T are presented in the Supplemental Material \cite{Sup}}.}
\label{fig:TE}
\end{figure}
For  $B\le4$\,T,  $\alpha_{\parallel}$ shows a discontinuity at $T_{\mathrm{Q}}\approx 0.11$\,K, indicating the onset of AFQ order. This result is in line with previous specific heat and electrical resistivity measurements\cite{Onimaru11}. At $B_{\mathrm{c}} \approx 5$\,T, the AFQ order is suppressed and a maximum, with a strongly enhanced value of $\alpha_{\parallel}\approx 43 \cdot 10^{-6}$\,K$^{-1}$, emerges. For higher magnetic fields the maximum broadens and shifts to higher temperatures. A distinct feature is the magnetic field induced divergence of $\alpha_{\parallel}$ for $T\rightarrow 0$ within the AFQ ordered phase.  A nuclear Schottky contribution is excluded as the cause, since the divergence vanishes for $B\ge5$\,T, which is at odds with the monotonic increase of a nuclear contribution as a function of magnetic field. 

In order to calculate the volume thermal expansion coefficient ($\beta=\alpha_{\parallel}+2\alpha_{\perp}$), it is necessary to determine both the longitudinal and transverse thermal expansion coefficients ($\alpha_{\parallel}$ and $\alpha_{\perp}$) for $B\parallel[001]$. A comparison of $\alpha_{\parallel}$ and $\alpha_{\perp}$ for intermediate magnetic fields is presented in Fig. 2\,a. It shows that $\alpha_{\perp}$ is the mirror image of $\alpha_{\parallel}$ with roughly half of its magnitude. The resulting volume thermal expansion $\beta$, which is approximately one order of magnitude smaller than the linear thermal expansion, is shown in Fig. 2\,b.  The relatively small volume changes around $B_{\mathrm{c}}\approx5$\,T indicate that the possible ferrohastatic order is not associated with a noticeable change in hybridization. This observation is distinct from magnetic Kondo lattice materials, which display large volume thermal expansion due to the high hydrostatic pressure sensitivity of the Kondo temperature \citep{Gegenwart16}. Prime examples are CeCu$_6$ and CeRu$_2$Si$_2$, which exhibit a volume expansion of $\beta=10^{-5}$\,K$^{-1}$ below 5\,K\cite{DeVisser89, DeVisser88}. This is one order of magnitude larger than the value observed for PrIr$_2$Zn$_{20}$.

In the following, we compare the experimental results for $\alpha_{\parallel}$ and $\alpha_{\perp}$ with a mean-field calculation which is based on a two sub-lattice model. The hamiltonian is given as
\begin{equation}
\begin{split}
\mathcal{H}_{\mathrm{A(B)}} = &\mathcal{H}_{\mathrm{CEF}}-g_{J}\mu_{\mathrm{B}}\boldsymbol{J}\boldsymbol{H}-g_{\varGamma_{3}}\left[O^0_2 \epsilon_{u}+O^2_2 \epsilon_{v} \right]\\
& -K_{\varGamma_3}\left[  O^0_2 \left<O^0_2\right>_{\mathrm{B(A)}}+O^2_2 \left<O^2_2\right> _{\mathrm{B(A)}} \right ]\\
& -K\boldsymbol{J}\left<\boldsymbol{J}\right>_{\mathrm{B(A)}},
\end{split}
\end{equation}
where $g_{J}$ is the Landé factor, $\mu_{\mathrm{B}}$ the Bohr magneton, $g_{\varGamma_{3}}$ the quadrupole-strain coupling constant, $K_{\varGamma_3}$ the interaction coefficient between $\varGamma_3$-type quadrupoles and $K$ the magnetic interaction coefficient. $\epsilon_{u}=(2\epsilon_{zz}-\epsilon_{xx}-\epsilon_{yy})/\sqrt{3}$ and $\epsilon_{v}= \epsilon _{xx} - \epsilon_{yy}$ denote the $\varGamma_3$ symmetry strains. The relative length change $\Delta L/L$ is proportional to the strain 
\begin{equation}
\epsilon_{\varGamma_{3}} = \frac{N g_{\varGamma_{3}}}{C^0_{\varGamma_{3}}}\left< O_{\varGamma_{3}}\right>,
\end{equation}
where $N=2.751 \cdot 10^{27}$\,1/m$^3$ is the number of Pr-ions per unit volume, $C^0_{{\varGamma}_{3}}=50.74$\,GPa the elastic modulus and $\left< O_{{\varGamma}_{3}} \right>$ the thermal average of the respective Stevens operator \cite{Ishii11}. The CEF effect is described by the Hamiltonian 
\begin{equation}
\mathcal{H_{\mathrm{CEF}}}=W\left[x\frac{O_4^0 + 5O^4_4}{60}+(1-|x|)\frac{O_6^0-21O_6^4}{1260}\right],
\end{equation}
where $W=-1.22$\,K and $x=0.537$ \cite{Lea62, Iwasa13}. 

The longitudinal and transverse relative length changes for magnetic fields $ \boldsymbol{B} \parallel [001]$ are estimated by
\begin{equation}
\frac{\Delta L}{L} \bigg| _{[001]}=\frac{1}{3}\epsilon_{\mathrm{B}}+\frac{1}{\sqrt{3}}\epsilon_{u},
\end{equation}
\begin{equation}
\frac{\Delta L}{L} \bigg| _{[100]}=\frac{1}{3}\epsilon_{\mathrm{B}}-\frac{1}{2 \sqrt{3}}\epsilon_{u}+\frac{1}{2}\epsilon_{v},
\end{equation}
where $\epsilon_{\mathrm{B}}=\epsilon_{xx}+\epsilon_{yy}+\epsilon_{zz}$ is the isotropic volume thermal expansion of the $\varGamma_1$ symmetry. In zero magnetic field, a tetragonal distortion does not exist and the linear thermal expansion corresponds to $\epsilon_{\mathrm{B}}/3$. As Fig. 1 illustrates, the zero field linear thermal expansion is vanishingly small as compared to the linear thermal expansion in magnetic fields $B\ge4$\,T. Thus, $\epsilon_{\mathrm{B}}$ is not included in the calculation, just as $\epsilon_{v}$, which is insignificantly small as compared to $\epsilon_{u}$. The quadrupole interaction is set to $K_{\varGamma_3}=-0.0067$\,K in order to reproduce the experimentally observed AFQ ordering temperature of $T_{\mathrm{Q}}=0.11$\,K in zero magnetic field. Both, the magnetoelastic constant ($g_{\varGamma_3}=-38$\,K) and the magnetic interaction ($K=-0.19$\,K) are determined by fitting the experimental data at high magnetic fields. 
\begin{figure}[htbp]
\includegraphics[width=\linewidth,keepaspectratio]{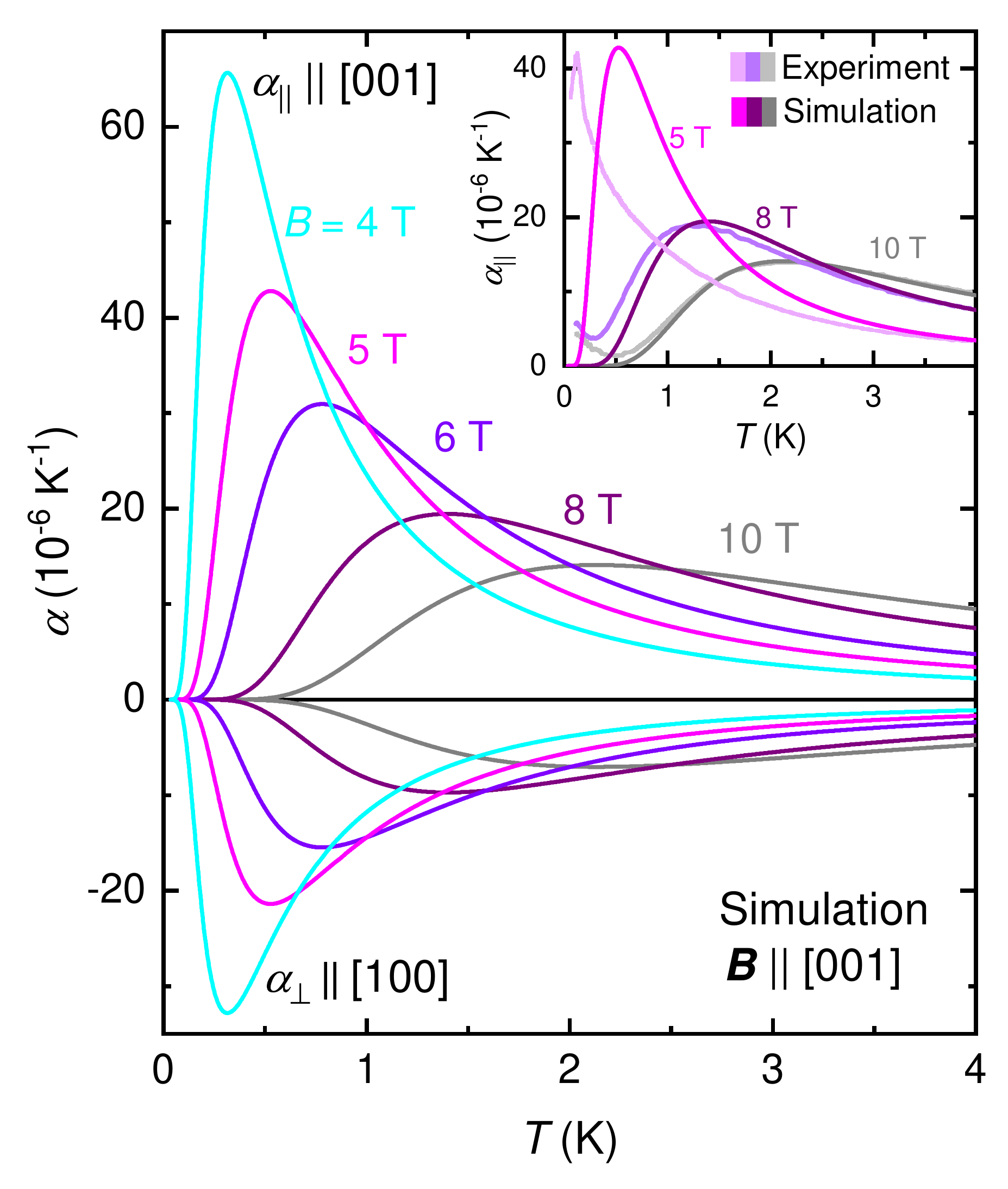}
\caption{Mean-field simulation of the longitudinal and transverse thermal expansion coefficients ($\alpha_{\parallel}$ and $\alpha_{\perp}$) for magnetic fields $\boldsymbol{B}\parallel [001]$. The inset shows a comparison of mean-field simulation and experimental data for selected magnetic fields.}
\label{fig:TE}
\end{figure}
Those values are comparable to the ones which were calculated by the elastic constants measurement ($|g_{\varGamma_3}|=30.9$\,K) and by the paramagnetic Curie temperature ($K=-0.35$\,K) \citep{Ishii11}. 

The simulation results are shown in Fig. 3, where the inset gives a direct comparison between experimental data and simulation. The opposite signs of  $\alpha_{\parallel}$ and $\alpha_{\perp}$ match with the experimental data and at the highest magnetic field of 10\,T, temperature dependence and absolute value fit very well. For magnetic fields close to $B_{\mathrm{c}}$, the experimentally determined extrema appear at much lower temperatures than the simulated ones. Moreover, the divergent behavior of $\alpha$ within the AFQ ordered state cannot be explained by the mean-field calculation. On approaching absolute zero temperature, entropy should go to zero as described by the third law of thermodynamics. Therefore, the linear thermal expansion coefficient $\alpha$, which measures the initial uniaxial stress derivative of entropy, should vanish as well \cite{Gegenwart16}. Only if entropy remains finite at absolute zero temperature, $\alpha$ can remain finite. Thus, the large value of $\alpha$ is a possible indication of residual entropy with a high sensitivity to uniaxial stress. This result is supported by specific heat measurements, which suggest a finite residual magnetic entropy for magnetic fields $1\,\mathrm{T} \le B \le {5\,\mathrm{T}}$, as well \cite{Onimaru16}.

Next, we turn to the linear and volume magnetostriction of PrIr$_{2}$Zn$_{20}$. Fig. 4\,a shows the longitudinal and transverse magnetostriction coefficients $(\lambda_{\parallel}$ and $\lambda_{\perp})$ for magnetic fields $\boldsymbol{B}\parallel[001]$ at different temperatures.
\begin{figure}
\includegraphics[width=\linewidth,keepaspectratio]{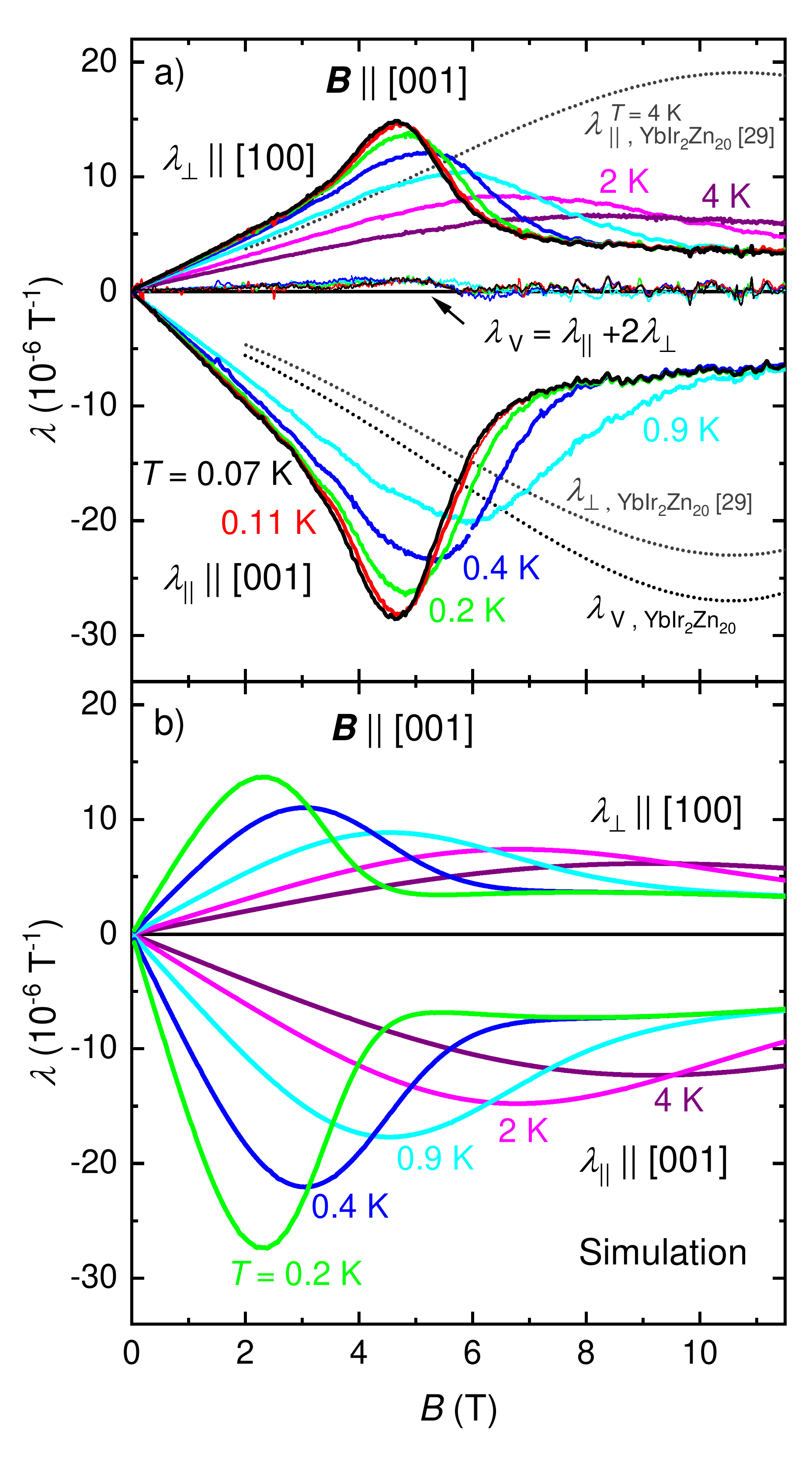}
\caption{ (a) Magnetic field dependence of the longitudinal and transverse magnetostriction coefficients ($\lambda_{\parallel}$ and $\lambda_{\perp}$) for \hbox{$\boldsymbol{B}\parallel [001]$} at various temperatures. The volume magnetostriction is determined as $\lambda_{\mathrm{V}}=\lambda_{\parallel}+2\lambda_{\perp}$. Additionally, linear and volume magnetostriction of YbIr$_2$Zn$_{20}$, which were extracted from Ref.\,29, are shown. (b) Mean-field simulation for $\lambda_{\parallel}$ and $\lambda_{\perp}$ at various temperatures}
\end{figure}
$\lambda_{\parallel}$ and $\lambda_{\perp}$ show opposite signs, whereby the contraction appears along and the expansion perpendicular to the magnetic field direction. At the lowest temperature of 0.07\,K, a sharp extremum appears at $B_{\mathrm{c}}\approx 4.7$\,T. This value coincides with the critical field of AFQ order \cite{Onimaru16}. With increasing temperature the extremum broadens and shifts to higher magnetic field.

In order to estimate the influence of the CEF effect, we performed the same mean-field simulation which was already used to calculate the linear thermal expansion coefficients. The simulation results are presented in \hbox{Fig. 4\,b.}
For $T\le0.9$\,K the experimentally determined extrema positions appear at much higher magnetic fields than the calculated ones, which points towards an additional contribution at intermediate magnetic fields and low temperatures. This mismatch was also observed for the linear thermal expansion and gives evidence that the behaviors close to $B_\mathrm{c}$ cannot be explained by the CEF effect.

Before we discuss the volume magnetostriction, we would like to address the effect of a possible misalignment between crystallographic [001] direction of the sample and magnetic field on $B_{\mathrm{c}}$. As $B_{\mathrm{c}}$ is highly anisotropic, with $B_{\mathrm{c}, { [001]}} < B_{\mathrm{c},  [110]} < B_{\mathrm{c},  [111]}$, such misalignment should cause an increase of $B_{\mathrm{c}}$. This error is likely more of an issue for the measurement perpendicular to magnetic field, where the exact alignment of the [001] direction is more complicated than for the measurement parallel to magnetic field. Since the linear coefficients display a sharp anomaly at $B_{\mathrm{c}}$, even a very little shift of $B_\mathrm{c}$ for $\lambda_{\perp}$ can cause a substantial error in the estimation of the volume magnetostriction \hbox{$\lambda_{\mathrm{V}}=\lambda_{\parallel}+2\lambda_{\perp}$}. To quantify this error, we compare $B_{\mathrm{c}}$ of $\lambda_{\parallel}$ and $\lambda_{\perp}$ at the lowest measured temperature of $0.07$\,K. It shows that $B_\mathrm{c, \lambda_{\parallel}}=0.968\,\cdot B_\mathrm{c, \lambda_{\perp}}$. Based on this result, the $B$-value of the $\lambda_{\perp}$ data presented in Fig. 4\,a is scaled by a constant factor 0.968. An explicit discussion of this error and the raw data can be found in the supplemental material.

Despite the huge uniaxial deformations, the volume of PrIr$_2$Zn$_{20}$ remains nearly constant in the whole magnetic field range. For comparison, linear and volume magnetostriction\cite{Takeuchi10} at $T=4$\,K of the isostructural material YbIr$_2$Zn$_{20}$ are additionally shown in Fig. 4\,a. The ground state of YbIr$_2$Zn$_{20}$ is a Kramers doublet which hybridizes with conduction electrons at low temperatures to form a heavy-Fermi liquid. The negative volume magnetostriction suggests the gradual increase of the Yb valence from a hybridized $\mathrm{Yb}^{2+}$ towards a $\mathrm{Yb}^{3+}$ state as a function of magnetic field \cite{Takeuchi10}.  By contrast, the volume magnetostriction of PrIr$_2$Zn$_{20}$ is vanishingly small, indicating that the valence of the Pr-ion remains nearly constant as a function of magnetic field.

Our study of thermal expansion and magnetostriction suggests that the high field phase of PrIr$_2$Zn$_{20}$ can be well described by a mean field CEF model. At intermediate magnetic fields and low temperatures, experiment and simulation show substantial differences confirming previous speculations on the formation of a new phase \cite{Onimaru16}. As already mentioned at the beginning of this article, ferrohastatic order which breaks the symmetry of the two equivalent channels of two-channel Kondo effect is a possible scenario \citep{Hoshino11}. Experimental signature of ferrohastatic order is a Fermi liquid state which was detected in the electrical resistivity. Recent theoretical work predicts that the Kondo temperature of this Fermi liquid state shows a very small magnetic field dependence \cite{VD2018}. In this case a rather field independent hybridization would be expected, compatible with the small volume magnetostriction.

To conclude, longitudinal and transverse thermal expansion and magnetostriction of PrIr$_2$Zn$_{20}$ for  $\boldsymbol{B}\parallel[001]$ are highly anisotropic. The respective volume changes are distinctly small as compared to Ce- and Yb-based Kramers Kondo lattice materials. This indicates that the previously observed peak in the field dependence of Seebeck coefficient \cite{Ikeura14} is not accompanied by a sizable change of c-f hybridization. Besides, a yet unexplainable divergence of linear thermal expansion inside the AFQ ordered phase was found.

We thank R. Küchler, A. Sakai and  C. Stingl for collaborative work and H. Kusunose, J. Otsuki and K. Hattori for helpful discussions. This work is financially supported by the German Science Foundation through project GE1640/8-1 and Grants-in-Aid from MEXT/JSPS of Japan, Nos.\,JP15KK0169, JP15H05886 (J-Physics), JP18H01182 and JP18KK0078.

\bibliography{ref}

\newpage

\onecolumngrid
\begin{center}
\textbf{\large Supplemental Material for ``Highly anisotropic strain dependencies in PrIr$_2$Zn$_{20}$"}
\end{center}
\section{Effect of Sample Misalignment}
In this section we discuss the effect of a misalignment between the crystallographic [001] direction and magnetic field on the critical field of antiferroquadrupolar order $B_{\mathrm{c}}$. Since $B_{\mathrm{c}}$ of PrIr$_2$Zn$_{20}$ is highly anisotropic, with $B_{\mathrm{c, [001]}}\approx5$\,T$<B_{\mathrm{c,[110]}}\approx10$\,T$<B_{\mathrm{c, [111]}}\approx12$\,T, a slight misalignment of the [001] direction should cause an increase of $B_{\mathrm{c}}$. 

Since longitudinal and transverse thermal expansion/magnetostriction were determined in two independent measurements it is important to compare the values of $B_{\mathrm{c}}$ for both measurement directions. In the measurement of longitudinal thermal expansion/magnetostriction the magnetic field is applied along the measurement direction of the relative length change. Therefore, simply the dilatometer itself has to be aligned parallel to magnetic field. The sample alignment in the transverse measurement is more difficult, as the sample has to be aligned carefully "by eye" and use of a special template inside the dilatometer at first. Afterwards, the dilatometer is rotated by 90 degrees. 
To reveal a possible sample misalignment Fig. 5 shows a comparison of the longitudinal and transverse magnetostriction coefficients ($\lambda_{\parallel}$ and $\lambda_{\perp}$), whereby $\lambda_{\perp}$ is multiplied by a factor -2 for a better comparability of the $B_\mathrm{c}$ values. It shows that $B_{\mathrm{c}}$ of $\lambda_{\perp}$ is slightly increased (approximately 3\%) as compared to  $\lambda_{\parallel}$. This suggests a small misalignment for the measurement of the transverse magnetostriction. Based on the geometry of the sample, the misalignment must be caused by a slight tilting from the crystallographic [001] direction towards the [011] direction with respect to magnetic field. To estimate the misalignment angle, we assume a linear field dependence of $B_{\mathrm{c}}$ between [001] and [011] direction. Then, the shift of the critical magnetic field corresponds to a misalignment angle of approximately 1.3 degrees.

\begin{figure}[h]
\includegraphics[width=0.59\textwidth,keepaspectratio]{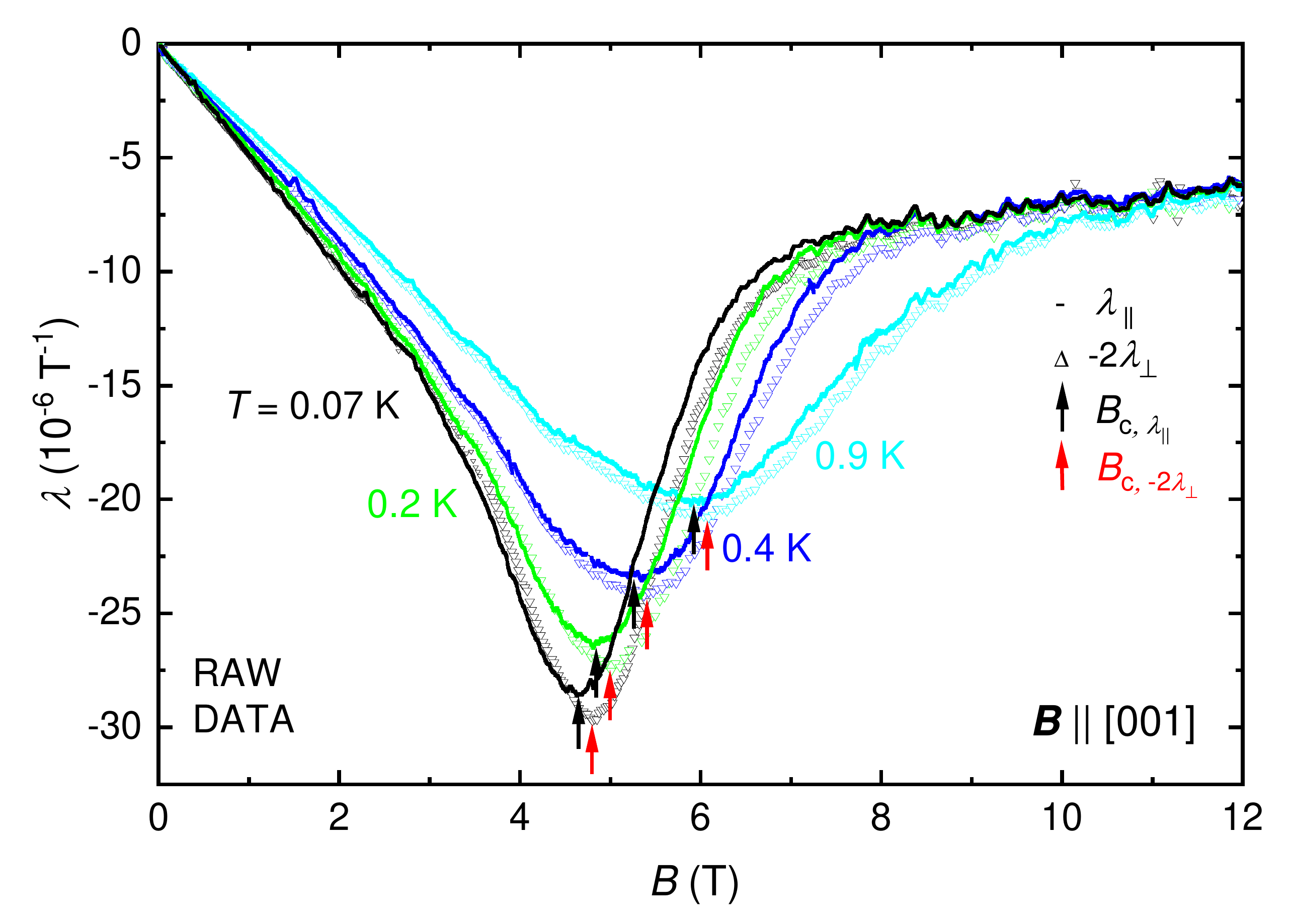}
\caption{Comparison of the critical magnetic field $B_\mathrm{c}$ for the longitudinal and transverse magnetostriction coefficients ($\lambda_\parallel$ and $\lambda_\perp$). For better comparability, the $\lambda_\perp$ data is multiplied by a factor -2. }
\label{fig:TE}
\end{figure}

Even though the effect of the misalignment is small it is of relevance for the calculation of the volume magnetostriction. At the lowest measured temperatures 0.07\,K and 0.11\,K, the longitudinal and transverse magnetostriction exhibit a sharp anomaly at $B_\mathrm{c}$ with opposite signs. Thus, already a very small shift of $B_\mathrm{c}$ in $\lambda_{\perp}$ results in an extrinsic anomaly in $\lambda_{\mathrm{V}}=\lambda_{\parallel}+2\lambda_\perp$. With increasing temperature, the extremum in the linear magnetostriction coefficients broadens and the misalignment effect is less critical.

To correct this error we scaled the $B$ value of the $\lambda_{\perp}$ data by a constant factor $B_\mathrm{c, \lambda_{\parallel}}/B_\mathrm{c, \lambda_{\perp}}=0.968$. This factor was determined at 0.07\,K, where the extremum is the sharpest. Fig. 6 shows the rescaled $\lambda_{\perp}$ data (multiplied by a factor -2 for better comparability) and the $\lambda_{\parallel}$ data. 
\begin{figure}[h]
\includegraphics[width=0.59\textwidth,keepaspectratio]{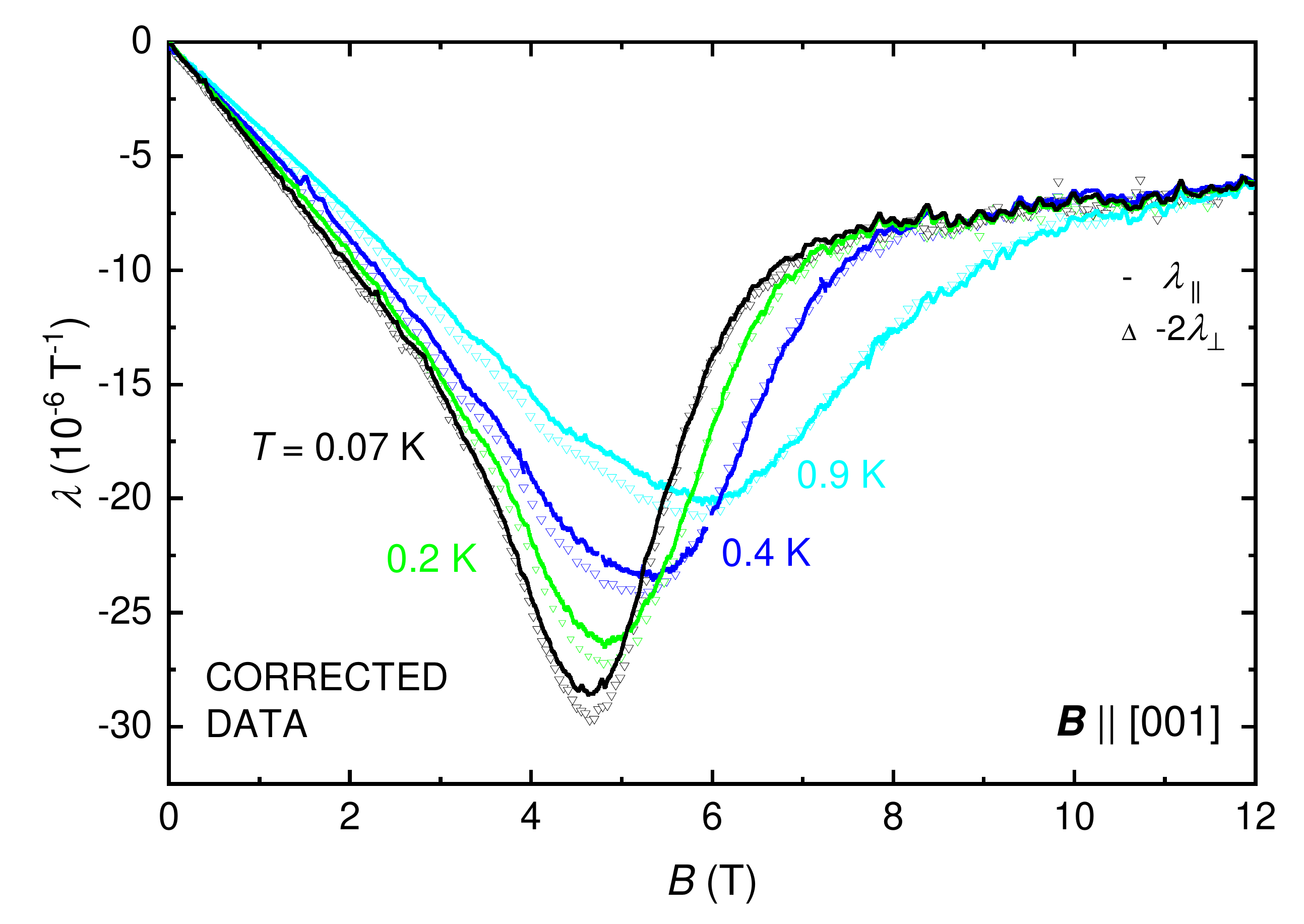}
\caption{Longitudinal and transverse magnetostriction coefficients ($\lambda_\parallel$ and $\lambda_\perp$). For better comparability, $\lambda_\perp$ is multiplied by a factor -2. The $B$ value of the $\lambda_{\perp}$ data is scaled by a factor 0.968 to correct a small misalignment in magnetic field.}
\label{fig:TE}
\end{figure}

Fig. 7 gives a comparison between $\lambda_{\mathrm{V}}$ calculated (a) from the raw $\lambda_\perp$ data and (b) from the corrected $\lambda_\perp$ data. The distinct feature around $B_\mathrm{c}$ in $\lambda_\mathrm{V}$ (Fig. 7\,a) shows that the slight shift of the critical field in the transverse measurement causes a significant error in the $\lambda_\mathrm{V}$ data. $\lambda_\mathrm{V}$ calculated from the corrected $\lambda_\perp$ data (Fig. 7\,b) is close to zero in the whole magnetic field range. Around $B_\mathrm{c}$ a very small enhancement of $\lambda_\mathrm{V}\approx 1\cdot 10^{-6}$\,T$^{-1}$ appears.
The little data variations at high magnetic fields are caused by quantum oscillations.
\begin{figure} 
    \subfigure{\includegraphics[width=0.45\textwidth]{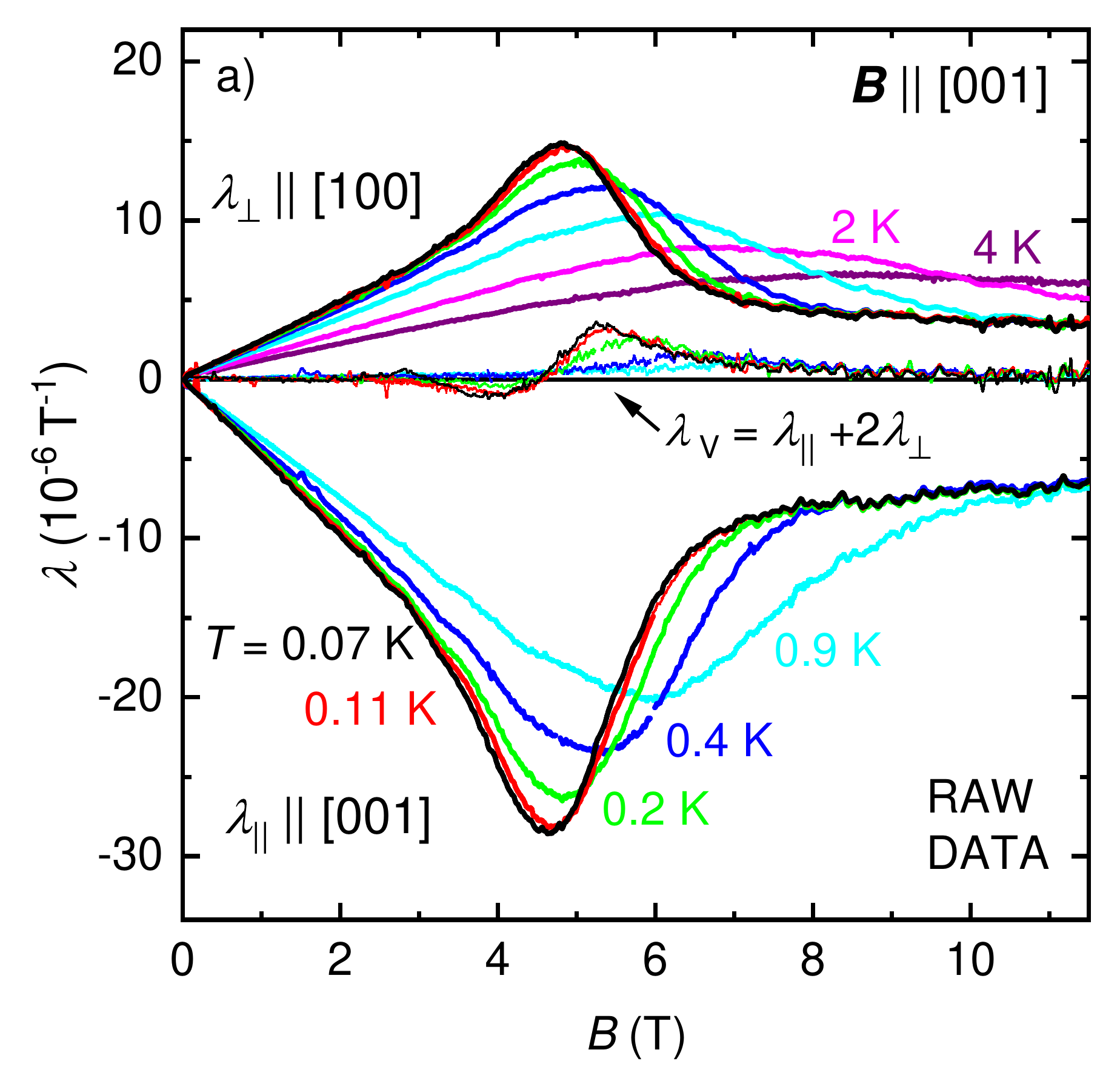}} 
    \hspace{4mm}
    \subfigure{\includegraphics[width=0.45\textwidth]{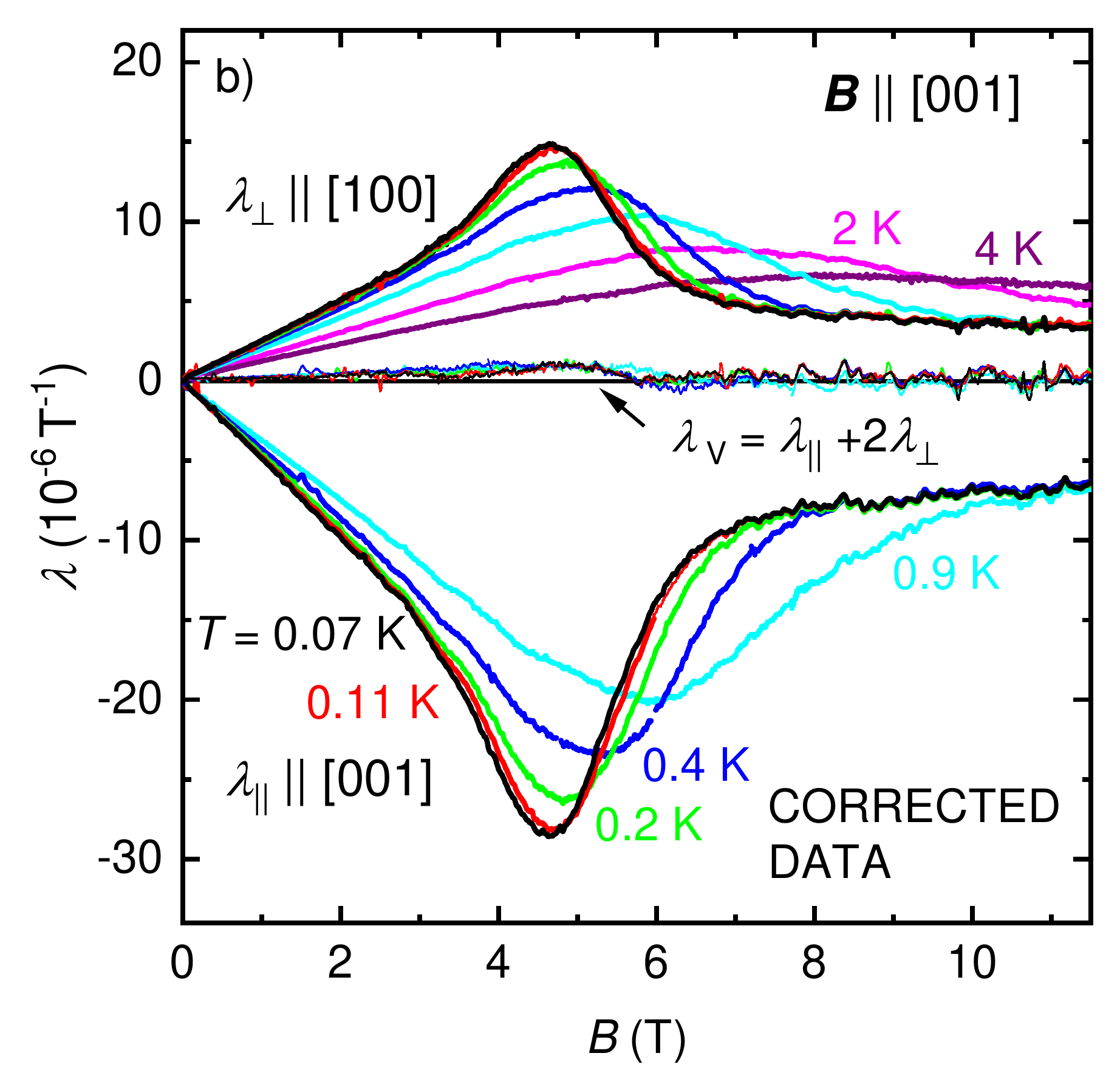}} 
\caption{Longitudinal $(\lambda_\parallel)$, transverse $(\lambda_\perp)$ and volume $(\lambda_\mathrm{V}=\lambda_\parallel+2\lambda_\perp)$ magnetostriction coefficients for (a) raw $\lambda_{\perp}$ data and (b) corrected $\lambda_{\perp}$ data. (a) $\lambda_\mathrm{V}$ shows an extrinsic anomaly around $B_{\mathrm{c}}$ at the lowest temperatures, due to the misalignment in the $\lambda_{\perp}$ measurement. (b) $\lambda_\mathrm{V}$, which is calculated by the corrected $\lambda_{\perp}$ data, is very small in the whole field range.} 
\end{figure} 

\section{Comparison between mean-field calculation and experiment at low magnetic fields}
In this section we present a comparison between experimental data and mean-field calculation for the longitudinal thermal expansion coefficient ($\alpha_{\parallel}$) in the range $B<4$\,T. As already mentioned in the main text, there are significant differences to the experiment for $B\le4$\,T. The large deviations clearly show that a theoretical treatment which goes beyond the simple mean-field calculation is required to describe the system at low magnetic fields. A comparison between experimental data and calculation is shown in Fig. 8. For $B<3$\,T the calculation exhibits a large anomaly due to the antiferroquadrupolar ordering.
\begin{figure}[h]
\includegraphics[width=0.58\textwidth,keepaspectratio]{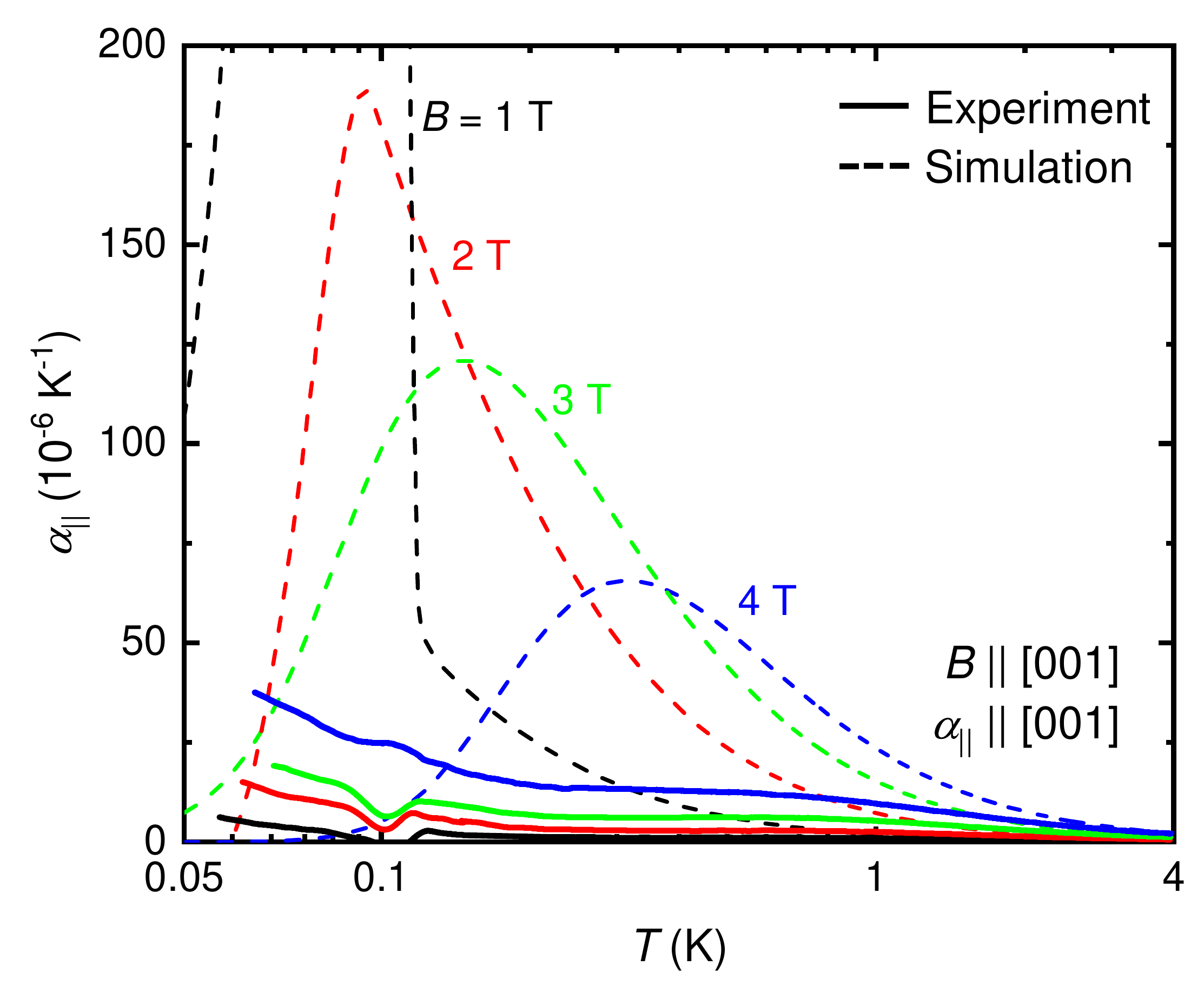}
\caption{Mean-field calculation (dashed lines) and experimental data (solid lines) for the longitudinal thermal expansion coefficient ($\alpha_{\parallel}$) for  magnetic fields $\boldsymbol{B}\parallel[001]$ in the range $B\le4$\,T.}
\label{fig:TE}
\end{figure}

\end{document}